%% file: SEH.tex
\documentclass{elsart}

\usepackage{cite}
\usepackage{graphicx}
\usepackage{dcolumn}
\usepackage{multirow}
\usepackage{braket}
\usepackage{caption}
\usepackage{subcaption}

\begin{document}

\begin{frontmatter}

\title{Highly accurate calculation of the resonances in the Stark effect in hydrogen}
\author{Francisco M Fern\'andez\thanksref{FMF}} and
\author{Javier Garcia}

\address{INIFTA (UNLP, CCT La Plata-CONICET), Divisi\'on Qu\'imica Te\'orica,
Blvd. 113 S/N,  Sucursal 4, Casilla de Correo 16, 1900 La Plata,
Argentina}

\thanks[FMF]{e--mail: fernande@quimica.unlp.edu.ar}

\begin{abstract}
We obtained accurate resonances for the Stark effect in hydrogen by means of
three independent methods. Two of them are based on complex rotation of the
coordinates and diagonalization of the Hamiltonian matrix (CRLM and CRCH).
The other one is based on the Riccati equations for the logarithmic
derivatives of factors of the wavefunction (RPM). The latter approach
enabled us to obtain the most accurate results and extremely sharp
resonances.
\end{abstract}

\end{frontmatter}

\section{ Introduction}

The Stark effect in hydrogen is an old problem in atomic spectroscopy and
one of the first triumphs of wave mechanics\cite{BS57,HMB74} (and references
therein). The Schr\"{o}dinger equation is separable in parabolic and squared
parabolic coordinates which facilitates the application of most approximate
methods\cite{HMB74}.

In a recent paper Fern\'{a}ndez-Menchero and Summers\cite{FS13} obtained the
complex eigenvalues and eigenfunctions of the Hamiltonian operator for the
hydrogen atom in a uniform electric field. They resorted to the
Lagrange-mesh basis set, proposed by Lin and Ho\cite{LH11} for the treatment
of the Yukawa potential in a uniform electric field, and the
complex-rotation method\cite{CHHRSW78}. They compared their results with
those obtained by Lin and Ho\cite{LH11}, Kolosov\cite{K87}, Rao and Li\cite
{RL95} and Ivanov\cite{I01} and overlooked the earlier impressive
calculations of Benassi and Grecchi\cite{BG80} and the accurate results
obtained by Fern\'{a}ndez\cite{F96b}. Benassi and Grecchi resorted to
complex scaling and a basis set of confluent hypergeometric functions that
is suitable when the Schr\"{o}dinger equation is written in squared
parabolic coordinates. On the other hand, Fern\'{a}ndez applied the
straightforward Riccati-Pad\'{e} method (RPM) that does not require the use
of complex coordinates.

The purpose of this paper is to calculate the Stark resonances as accurately
as possible by means of the methods proposed by Fern\'{a}ndez-Menchero and
Summers\cite{FS13}, Benassi and Grecchi\cite{BG80} and Fern\'{a}ndez\cite
{F96b} and compare the results with those obtained by the authors already
mentioned and also by Damburg and Kolosov\cite{DK76}. There is a vast
literature on the hydrogen atom in a uniform electric field but we restrict
present discussion to some of the available calculations that we deem are
suitable for comparison.

In section~\ref{sec:Stark} we outline the main ideas about separating the
Schr\"{o}dinger equation in parabolic and squared parabolic coordinates. In
sections \ref{sec:CRLM}, \ref{sec:CRCH} and \ref{sec:RPM} we briefly
introduce the methods of Fern\'{a}ndez-Menchero and Summers\cite{FS13},
Benassi and Grecchi\cite{BG80} and the RPM\cite{F96b}, respectively. In
section~\ref{sec:numerical} we compare the results of various approaches and
in section~\ref{sec:conclusions} we summarize the main results and draw
conclusions.

\section{Stark effect in hydrogen}

\label{sec:Stark}

The Schr\"{o}dinger equation in atomic units is
\begin{eqnarray}
H\psi &=&E\psi  \nonumber \\
H &=&-\frac{1}{2}\nabla ^{2}-\frac{1}{r}+Fz,  \label{eq:Schro_Stark}
\end{eqnarray}
where $F$ is the intensity of the uniform electric field assumed to be
directed along the $z$ axis.

This equation is separable in parabolic coordinates
\begin{eqnarray}
&&x =\sqrt{\xi \eta }\cos \phi ,\;y=\sqrt{\xi \eta }\sin \phi ,\;z=\frac{\xi
-\eta }{2}  \nonumber \\
&&\xi \geq 0,\;\eta \geq 0,\;0\leq \phi \leq 2\pi .  \label{eq:parabolic}
\end{eqnarray}
If we write
\begin{equation}
\psi (x,y,z)=(\xi \eta )^{-1/2}u(\xi )v(\eta )e^{im\phi },\;m=0,\pm 1,\pm
2,\ldots ,  \label{eq:psi_parabolic}
\end{equation}
then we obtain two equations of the form
\begin{equation}
\left( \frac{d^{2}}{dx^{2}}+\frac{1-m^{2}}{4x^{2}}+\frac{E}{2}-\sigma \frac{F%
}{4}x+\frac{A_{\sigma }}{x}\right) \Phi (x)=0,
\label{eq:separated_parabolic}
\end{equation}
where $\sigma =\pm 1$ and $A_{+}=A$ and $A_{-}=1-A$ are separation
constants. When $\sigma =1$, $x=\xi $ and $\Phi (\xi )=u(\xi )$; when $%
\sigma =-1$, $x=\eta $ and $\Phi (\eta )=v(\eta )$.

The Schr\"{o}dinger equation (\ref{eq:Schro_Stark}) is also separable in
squared parabolic coordinates

\begin{eqnarray}
&&x=\mu \nu \cos \phi ,\;y=\mu \nu \sin \phi ,\;z=\frac{\mu ^{2}-\nu ^{2}}{2}
\nonumber \\
&&\mu \geq 0,\;\nu \geq 0,\;0\leq \phi \leq 2\pi .  \label{eq:parabolic_sq}
\end{eqnarray}
If in this case we write
\begin{equation}
\psi (x,y,z)=(\mu \nu )^{-1/2}u(\mu )v(\nu )e^{im\phi },
\end{equation}
then we obtain two equations of the form
\begin{equation}
\left( \frac{d^{2}}{dx^{2}}+\frac{1-4m^{2}}{4x^{2}}+2Ex^{2}-\sigma
Fx^{4}+Z_{\sigma }\right) \Phi (x)=0,  \label{eq:separated_parabolic_sq}
\end{equation}
where, $\sigma =\pm 1$ and $Z_{+}=Z$ and $Z_{-}=4-Z$ are the separation
constants. When $\sigma =1$, $x=\mu $ and $\Phi (\mu )=u(\mu )$; when $%
\sigma =-1$, $x=\nu $ and $\Phi (\nu )=v(\nu )$.

The solutions to the equations in either set of coordinates are commonly
labelled by the quantum numbers $n_{1},n_{2}=0,1,2\ldots $ and $m=0,1,\ldots
$, and the notation $\left| n_{1},n_{2},m\right\rangle $ is suitable for
referring to them. We will sometimes resort to the principal quantum number $%
n=n_{1}+n_{2}+|m|+1$ to denote a set of states. Obviously, $m$ is the only
good quantum number; the other ones refer to the states of the hydrogen atom
and are valid when $F=0$.

\section{Complex rotation and Laguerre-mesh basis set}

\label{sec:CRLM}

Fern\'{a}ndez-Menchero and Summers\cite{FS13} decided to treat the
Schr\"{o}dinger equation as nonseparable. The Hamiltonian operator in
parabolic coordinates reads
\begin{equation}
H=-\frac{2}{\xi +\eta }\left[ \frac{\partial }{\partial \xi }\left( \xi
\frac{\partial }{\partial \xi }\right) +\frac{\partial }{\partial \eta }%
\left( \eta \frac{\partial }{\partial \eta }\right) \right] -\frac{1}{2\,\xi
\eta }\frac{\partial ^{2}}{\partial \phi ^{2}}-\frac{2}{\xi +\eta }+F\frac{%
\xi -\eta }{2},  \label{eq:H_parabolic}
\end{equation}
and the authors proposed the variational ansatz
\begin{eqnarray}
\psi \left( \xi ,\eta ,\phi \right) &=&\frac{1}{\sqrt{2\pi }}e^{im\phi
}\sum_{k=1}^{N}\sum_{l=1}^{N}c_{klm}e^{-\frac{\xi +\eta }{2}}\left( \xi \eta
\right) ^{\frac{|m|}{2}}\Lambda _{Nk}(\xi )\Lambda _{Nl}(\eta )
\label{eq:ansatz_FS} \\
\Lambda _{Nk}(x) &=&(-1)^{k}\sqrt{x_{k}}\frac{L_{N}(x)}{x-x_{k}},
\end{eqnarray}
where $L_{N}(x)$ is the Laguerre polynomial of degree $N$ and $x_{k}$ its $k$%
-th zero. In order to obtain the resonances they resorted to the well-known
complex rotation method\cite{CHHRSW78} that in this case is given by the
transformation $(\xi ,\eta )\rightarrow (e^{i\vartheta }\xi ,e^{i\vartheta
}\eta )$, where $\vartheta$ is the rotation angle. The eigenvalues and
expansion coefficients are given by the secular equation
\begin{equation}
(\mathbf{H}-E\mathbf{S})\mathbf{C}=0,  \label{eq:secular}
\end{equation}
where the elements of the $N^{2}\times N^{2}$ matrices $\mathbf{H}$ and $%
\mathbf{S}$ are explicitly shown elsewhere\cite{FS13} and the elements of
the column vector $\mathbf{C}$ are the coefficients $c_{klm}$. Note that the
integrals appearing in the matrix elements of both $\mathbf{H}$ and $\mathbf{%
S}$ should be calculated numerically and when we increase $N$ we have to
calculate all those integrals again. For brevity we will call this method
CRLM.

\section{Complex scaling and confluent hypergeometric basis set}

\label{sec:CRCH}

In order to obtain the resonances Benassi and Grecchi\cite{BG80} resorted to
equation (\ref{eq:separated_parabolic_sq}) and a basis set of the form
\begin{equation}
\varphi _{m,n}(x)=\frac{2(m+n)!}{m!n!}e^{-x^{2}/2}x^{m+\frac{1}{2}%
}F(-n,m+1;x^{2}),  \label{eq:basis_BG}
\end{equation}
where $F(a,b,z)$ is the confluent hypergeometric function. In this case the
authors resorted to the complex scaling method that is based on the
transformation $(\mu ,\nu )\rightarrow (\lambda ^{1/2}\mu ,\lambda ^{1/2}\nu
)$, where $\lambda $ is a complex number. The complex scaling method
contains the complex rotation method as a particular case because $\lambda
=|\lambda |e^{i\vartheta }$ and the proper choice of $|\lambda |$ enables
one to improve the convergence of the approach.

In this case all the elements of the relevant pentadiagonal matrix can be
calculated analytically and are independent of the matrix dimension. This
approach will be called CRCH from now on.

\section{The Riccati-Pad\'{e} method}

\label{sec:RPM}

We can apply the RPM to the eigenvalue equations derived in either parabolic
or squared parabolic coordinates. In the earlier application of the approach
Fern\'{a}ndez\cite{F96b} chose the former and here we resort to the latter.
It is worth mentioning that the performance of the RPM in both sets of
coordinates is identical and that the reason for the selection of the
squared parabolic coordinates is to have a closer contact between the RPM
and the CRCH method of Benassi and Grecchi\cite{BG80}. The regularized
logarithmic derivative
\begin{equation}
f(x)=\frac{s}{x}-\frac{\Phi ^{\prime }(x)}{\Phi (x)},\;s=|m|+\frac{1}{2},
\label{eq:f(x)}
\end{equation}
can be expanded in a Taylor series
\begin{equation}
f(x)=\sum_{j=0}^{\infty }f_{j}x^{2j+1},  \label{eq:f(x)_exp}
\end{equation}
where the coefficients $f_{j}$ are polynomial functions of $E$ and $Z$. The
details of the method are outlined elsewhere\cite{F96b}; here it suffices to
say that we construct Hankel determinants of the form
\begin{equation}
H_{D}^{d}(E,Z,F)=\left|
\begin{array}{cccc}
f_{d+1} & f_{d+2} & \ldots & f_{D+d} \\
f_{d+2} & f_{d+3} & \ldots & f_{D+d+1} \\
&  & \ddots &  \\
f_{D+d} & f_{D+d+1} & \ldots & f_{2D+d-1}
\end{array}
\right| ,  \label{eq:Hankel}
\end{equation}
and obtain the approximate eigenvalues $E^{[D,d]}$ from the roots of the set
of nonlinear equations
\begin{equation}
H_{D}^{d}(E,Z,F)=H_{D}^{d}(E,4-Z,-F)=0.  \label{eq:RPM_quantization}
\end{equation}

The main advantage of the RPM is the enormous rate of convergence which
enables us to obtain very accurate eigenvalues with determinants of
relatively small dimension. However, the great number of roots in the
neighborhood of each eigenvalue makes it difficult to find the optimal
sequence that converges to it. Since we resort to the Newton-Raphson
algorithm to obtain the roots of the system of equations (\ref
{eq:RPM_quantization}) we have to choose the starting point quite close to
the chosen root. We will discuss this point briefly in Section~\ref
{sec:numerical}. The RPM is most suitable for the treatment of separable
problems.

\section{Numerical calculations}

\label{sec:numerical}

In order to apply the CRLM\cite{FS13} we calculated all the integrals
numerically with a tolerance of $10^{-15}$. For each value of $F$, $m$ and $%
N $ we varied the rotation angle $\vartheta $ between $0.3$ and $0.7$
looking for those eigenvalues that remained almost constant. We could
reproduce the results in the literature with matrices of dimension $N=30$%
\cite{FG15}.

The only exact quantum number is $m$, however, it is customary to resort to
the quantum numbers of the isolated hydrogen atom in order to label the
energies and states of the Stark problem. Some authors choose the parabolic
quantum numbers $n_{1},n_{2}=0,1,2,\ldots $\cite{BG80} and others the
principal quantum number $n=n_{1}+n_{2}+|m|+1$ and $k=n_{1}-n_{2}$\cite{FS13}%
.

In the case of CRCH we first solved the eigenvalue equations that yield the
eigenvalues $Z_{+}$ and $Z_{-}$ and then applied the Newton-Raphson method
to solve the equation $Z_{+}+Z_{-}-4=0$. Details of the calculation are
given elsewhere\cite{BG80}; we just mention that in order to obtain a
starting point for the Newton-Raphson method we resorted to the results
provided by perturbation theory\cite{F00}.

The Hankel determinants that appear in the RPM are polynomial functions of $%
E $ and $Z$ of great degree. For this reason it is necessary to handle
complex numbers with sufficiently great precision and we resorted to the GNU
MPC library\cite{EGTZ15}. The Hankel determinants can be calculated
numerically by means of the well known recurrence relation

\begin{equation}
H_{D}^{d}=\frac{H_{D-1}^{d}H_{D-1}^{d+2}-\left( H_{D-1}^{d+1}\right) ^{2}}{%
H_{D-2}^{d+2}}  \label{eq:Hankel_recurrence}
\end{equation}
with the initial conditions $H_{0}^{d}=1$ and $H_{1}^{d}=f_{d+1}$. Once we
calculate the desired determinants we obtain the eigenvalue and separation
constant by means of the Newton-Raphson method. In order to have a suitable
starting point we resorted to CRCH results.

\begin{figure}[H]
\caption{Convergence of the RPM resonances for $F=0.005$}
\label{fig:conv2}
\begin{center}
\begin{subfigure}[b]{0.48\textwidth}
\resizebox{6cm}{!}{\input{conv2_Re}}
\end{subfigure}
\begin{subfigure}[b]{0.48\textwidth}
\resizebox{6cm}{!}{\input{conv2_Im}}
\end{subfigure}
\end{center}
\end{figure}
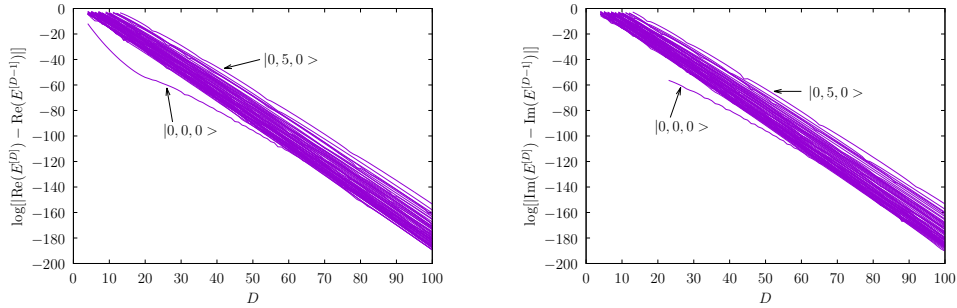

The remarkable rate of convergence of the RPM is clearly illustrated by the
calculation of the logarithmic error $\log \left| \alpha ^{[D]}-\alpha
^{[D-1]}\right| $ where $\alpha $ stands for either $\mathrm{Re}E$ or $%
\mathrm{Im}E$. We do not indicate the value of $d$ explicitly because in
present calculations we have chosen $d=0$. Figures \ref{fig:conv1} and \ref
{fig:conv2} show the logarithmic error for all the resonances with $%
n=1,2,\ldots ,6$ and field strengths $F=0.001$ and $F=0.005$, respectively.
As the quantum numbers increase the minimum value of $D$ at which the
resonance appears also increases. For example, the lowest resonance appears
as a root of the Hankel determinant with $D=2$ and $\left|
0,5,0\right\rangle $ appears at $D=11$. The rate of convergence is greater
when the root of the Hankel determinant is real. This fact is clearly shown
in Fig.~\ref{fig:conv2} where the imaginary part of $\left|
0,0,0\right\rangle $ appears at $D=22$. When $F=0.001$ the imaginary part of
the lowest resonance appears at $D=103$ which is the reason why the rate of
convergence for this resonance is considerably greater than the other ones
for all $D\leq 100$. When $D>103$ the rate of convergence for the lowest
resonance becomes similar to the other ones.

The resonances explicitly labelled in figures \ref{fig:conv1} and \ref
{fig:conv2} are shown in Table~\ref{tab:RPM_1} with their number of digits
truncated to a reasonable size.

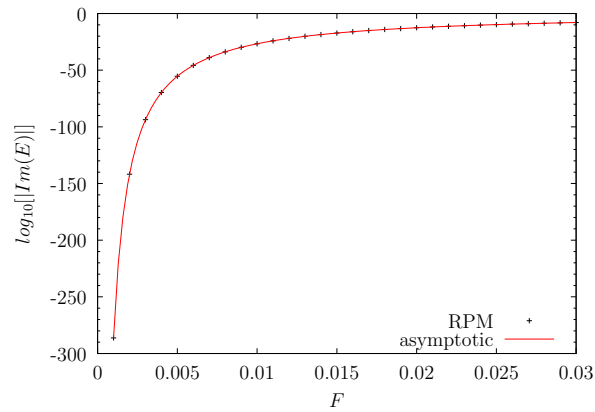
\begin{figure}[H]
\begin{center}
\resizebox{8cm}{!}{\input{RPM_vs_asinto}}
\end{center}
\caption{Width of the lowest resonance calculated by means of the RPM
(circles) and the asymptotic expansion (\ref{eq:Gamma_asymp}) (line)}
\label{fig:Gammas}
\end{figure}

Figure~\ref{fig:Gammas} shows that our estimated value of
$\mathrm{Im}E$ for the lowest resonance is in perfect agreement
with the analytic asymptotic formula derived by Benassi and
Grecchi\cite{BG80}:
\begin{equation}
\left| \mathrm{Im}E\right| \sim 2F^{-1}e^{-2/(3F)}\left( 1-8.91\bar{6}%
F+25.57F^{2}+O(F^{3})\right) .  \label{eq:Gamma_asymp}
\end{equation}

\begin{figure}[H]
\caption{Real and imaginary parts of the resonance $\Ket{39,0,0}$: (a)
\protect\cite{K87}, (b) RPM}
\label{fig:n=40}
\begin{center}
\begin{subfigure}[b]{0.48\textwidth}
\resizebox{6cm}{!}{\input{39-0-0_Re}}
\end{subfigure}
\begin{subfigure}[b]{0.48\textwidth}
\resizebox{6cm}{!}{\input{39-0-0_Im}}
\end{subfigure}
\end{center}
\end{figure}
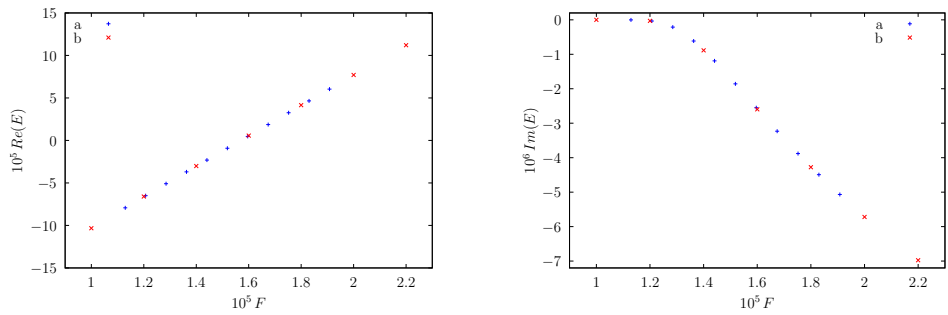

The RPM is also suitable for the calculation of higher resonances. For
example, tables \ref{tab:040} and \ref{tab:400} compare present results
obtained by the CRCH and RPM for two states with $n=5$ with those obtained
earlier by Damburg and Kolosov\cite{DK76}. Table~\ref{tab:n=10} compares
present RPM results for some states with $n=10$ with those obtained by
Kolosov\cite{K87}. The discrepancy in the imaginary part for the case $%
\Ket{0,9,0}$, $F=2.2\times 10^{-5}$, is probably due to a misprint in that
reference. Table~\ref{tab:n=40} shows the resonance $\Ket{39,0,0}$ for
several values of the field strength. We do not compare these results with
those of Kolosov\cite{K87} because he did not indicate the conversion factor
from atomic units to $Vcm^{-1}$ shown in his table. However, Fig.~\ref
{fig:n=40} shows that both sets of results are in reasonable agreement.

In the tables discussed above we have truncated present RPM
results to a reasonable number of digits. We have obtained them
with much higher accuracy as suggested by figures \ref{fig:conv1}
and \ref{fig:conv2}. For example, for the lowest resonance and
field strength $F=0.005$ we obtained

\begin{eqnarray}
\mathrm{Re}E
&=&-0.50005628479379296933177394769143288196325092731889137262135731
\nonumber \\
&&28725736315548994436307340293823812601699152241599625041068943791
\nonumber \\
&&42099665225189334039046848974164185728077545219665133771938893895
\nonumber \\
&&64251327341968732189236225621425838831553440690618168917215735013
\nonumber \\
&&803880912033036  \nonumber \\
\mathrm{Im}E
&=&-4.74901370837102040886757127120827250441845432417751748825418970
\nonumber \\
&&22400488040285011762035775189238632536585799373474503067879411046
\nonumber \\
&&22147574080708907330396144467615023762954201754322979890803189455
\nonumber \\
&&51562966634796276868224\times 10^{-56}
\end{eqnarray}
with Hankel determinants of dimension $D\leq 150$. In principle we expect
that a properly truncated perturbation series will exhibit an accuracy of
the order of $\left| \mathrm{Im}E\right| $. On summing the first 130 terms
of the perturbation series calculated by means of the hypervirial
perturbative method\cite{F00} we obtained the following result:
\begin{equation}
E^{PT}=-0.5000562847937929693317739476914328819632509273188913726,
\end{equation}
that agrees with the RPM one to the last digit. It is not easy to
obtain such a sharp resonance by means of other approaches; for
example, Fern\'{a}ndez-Menchero and Summers\cite{FS13} estimated
$\mathrm{Re}E=-0.5000553416$ and $\mathrm{Im}E=0.8944475605\times
10^{-7}$. We calculated the real part of this resonance more
accurately by means of the CRLM but were unable to obtain a
reasonable estimate of the imaginary part\cite{FG15}.

The RPM enables one to calculate even sharper resonances; for
example, for the lowest one and $F=0.001$ we obtained
\begin{eqnarray}
\mathrm{Re}E
&=&-0.50000225005555178356591589970608204532866714376652965654995937
\nonumber \\
&&97019283545891048870035463753481536961447568150634794700138591827
\nonumber \\
&&91549628581187487453336046428670620173909589867079695807271725700
\nonumber \\
&&47474205292728633151353049600188535220623998127315129221076077663
\nonumber \\
&&756392409425470889188167975544640438386213612059475282765271923  \nonumber
\\
\mathrm{Im}E &=&-5.854592875137598393486482622915575\times 10^{-287}
\end{eqnarray}
with Hankel determinants of dimension $D\leq 130$. In this case perturbation
theory of order 600 (300 nonzero terms) yields

\begin{eqnarray}
E^{PT}
&=&-0.50000225005555178356591589970608204532866714376652965654995937970
\nonumber \\
&&19283545891048870035463753481536961447568150634794700138591827915496
\nonumber \\
&&28581187487453336046428670620173909589867079695807271725700474742052
\nonumber \\
&&92728633151353049600188535220623998127315129221076077663756392409425
\nonumber \\
&&47088918816797549
\end{eqnarray}

\section{Conclusions}

\label{sec:conclusions}

We have calculated the resonances of the Stark effect in hydrogen by means
of three independent methods. Although we were able to improve the CRLM
calculation considerably\cite{FG15} we think that the CRCH is far more
efficient. However, the RPM yielded considerable more accurate results and
enabled us to obtain extremely sharp resonances that we were not able to
obtain by means of the other two methods. The reason is that the accuracy of
the real part should be at least of the order of the imaginary one. We were
able to attain such an accuracy in the calculation of the roots of the RPM
equations (\ref{eq:RPM_quantization}) thanks to the GNU MPC library\cite
{EGTZ15}. We think that it is almost impossible to do the same by means of
the CRLM because of the numerical calculation of the matrix elements. In
principle, one can obtain the resonances with any degree of accuracy by
means of the CRCH but such calculation would require a great deal of
ingenuity. For this reason we think that the RPM is an extremely suitable
benchmark to test other approaches on separable models.

\begin{table}[H]
\caption{Resonances for the states appearing in figures \ref{fig:conv1} and
\ref{fig:conv2}}
\label{tab:RPM_1}\centering
\begin{tabular}{|l|l|l|l|}
\hline
\multicolumn{1}{|c}{$F$} & \multicolumn{1}{|c}{Resonance} &
\multicolumn{1}{|c}{Re($E$)} & \multicolumn{1}{|c|}{Im($E$)} \\ \hline
\multirow{5}{*}{0.001} & $\Ket{0,0,0}$ & $-0.5000022500555518$ & $%
-6.584169959231863\times 10^{-287}$ \\
& $\Ket{0,0,1}$ & $-0.1250782240371032$ & $-8.433615180808857\times 10^{-33}$
\\
& $\Ket{0,1,0}$ & $-0.1280858350607099$ & $-2.060525710039887\times 10^{-31}$
\\
& $\Ket{1,0,0}$ & $-0.1220826861326878$ & $-3.395926205766083\times 10^{-34}$
\\
& $\Ket{0,5,0}$ & $-0.05215538955477732$ & $-2.594493723108199\times 10^{-2}$
\\ \hline
\multirow{5}{*}{0.005} & $\Ket{0,0,0}$ & $-0.5000562847937930$ & $%
-4.749013708371020\times 10^{-56}$ \\
& $\Ket{0,0,1}$ & $-0.1271466127039709$ & $-1.307642723230557\times 10^{-5}$
\\
& $\Ket{0,1,0}$ & $-0.1426186075727077$ & $-5.297223183652474\times 10^{-5}$
\\
& $\Ket{1,0,0}$ & $-0.1120619240019938$ & $-2.864684219868783\times 10^{-6}$
\\
& $\Ket{0,5,0}$ & $-0.1213596730003857$ & $-1.176260968442979\times 10^{-1}$
\\ \hline
\end{tabular}
\end{table}

\begin{table}[H]
\caption{Resonance $\ket{0,4,0}$ from
reference~\protect\cite{DK76} (a) and present calculation by means
of CRCH (b) and RPM (c)} \label{tab:040}\scalebox{0.8}{
\begin{tabular}{|c|c|l|l|}
\hline
\multicolumn{1}{|c|}{$F$} &  & \multicolumn{1}{c|}{Re($E$)} &
\multicolumn{1}{c|}{Im($E$)} \\ \hline
\multirow{3}{*}{$0.00010$} & a & $-0.0231791962$ & $-2.1\times 10^{-12}$ \\
& b & $-0.02317919625030$ & $-2.1135\times 10^{-12}$ \\
& c & $-0.02317919625030518$ & $-2.113884073268850\times 10^{-12}$ \\ \hline
\multirow{3}{*}{$0.00015$} & a & $-0.024956749$ & $-9.595\times 10^{-7}$ \\
& b & $-0.024956750918078$ & $-9.6007202913\times 10^{-7}$ \\
& c & $-0.02495675091807878$ & $-9.600720291331372\times 10^{-7}$ \\ \hline
\multirow{3}{*}{$0.00020$} & a & $-0.02697136$ & $-8.9150\times 10^{-5}$ \\
& b & $-0.0269800814710915$ & $-9.36280360832384\times 10^{-5}$ \\
& c & $-0.02698008147109154$ & $-9.362803608323849\times 10^{-5}$ \\ \hline
\multirow{3}{*}{$0.00025$} & a & $-0.02896828$ & $-4.2655\times 10^{-4}$ \\
& b & $-0.02912946983310681$ & $-4.868994650393436\times 10^{-4}$ \\
& c & $-0.02912946983310681$ & $-4.868994650393436\times 10^{-4}$ \\ \hline
\multirow{3}{*}{$0.00030$} & a & $-0.0305381$ & $-9.849\times 10^{-4}$ \\
& b & $-0.03122955458572655$ & $-1.127494087615666\times 10^{-3}$ \\
& c & $-0.03122955458572655$ & $-1.127494087615666\times 10^{-3}$ \\ \hline
\multirow{3}{*}{$0.00035$} & a & $-0.0314338$ & $-1.8217\times 10^{-3}$ \\
& b & $-0.03323652729915596$ & $-1.945324601526496\times 10^{-3}$ \\
& c & $-0.03323652729915596$ & $-1.945324601526496\times 10^{-3}$ \\ \hline
\multirow{3}{*}{$0.00040$} & a & $-0.031408$ & $-3.17565\times 10^{-3}$ \\
& b & $-0.03512209724011620$ & $-2.892832253491630\times 10^{-3}$ \\
& c & $-0.03512209724011620$ & $-2.892832253491630\times 10^{-3}$ \\ \hline
\multirow{3}{*}{$0.00045$} & a & $-0.02998$ & $-6.365\times 10^{-3}$ \\
& b & $-0.03687445248862566$ & $-3.911655467856354\times 10^{-3}$ \\
& c & $-0.03687445248862566$ & $-3.911655467856354\times 10^{-3}$ \\ \hline
\end{tabular}

}

\end{table}

\begin{table}[H]
\caption{Resonances for the state $\ket{4,0,0}$ from
reference~\protect\cite {DK76} (a) and present calculation by
means of CRCH (b) (b) and RPM (c)} \label{tab:400}\scalebox{0.8}{
\begin{tabular}{|c|c|l|l|}
\hline
\multicolumn{1}{|c|}{$F$} &  & \multicolumn{1}{c|}{Re($E$)} &
\multicolumn{1}{c|}{Im($E$)} \\ \hline
\multirow{3}{*}{$0.00015$} & a & $-0.0158077645$ & $-1\times 10^{-11}$ \\
& b & $-0.01580776440749585$ & $-7.156147028941416\times 10^{-12}$ \\
& c & $-0.01580776440749585$ & $-7.156147028941416\times 10^{-12}$ \\ \hline
\multirow{3}{*}{$0.00020$} & a & $-0.0145352049$ & $-2.013{-8}$ \\
& b & $-0.01453520517676726$ & $-2.012419057345574\times 10^{-8}$ \\
& c & -0.01453520517676726 & $-2.012419057345574\times 10^{-8}$ \\ \hline
\multirow{3}{*}{$0.00025$} & a & $-0.013328925$ & $-1.63595\times 10^{-6}$
\\
& b & $-0.01332892813256598$ & $-1.637235677233378\times 10^{-6}$ \\
& c & $-0.01332892813256598$ & $-1.637235677233378\times 10^{-6}$ \\ \hline
\multirow{3}{*}{$0.00030$} & a & $-0.01220093$ & $-2.0833\times 10^{-5}$ \\
& b & $-0.01220135935615766$ & $-2.104916128829678\times 10^{-5}$ \\
& c & $-0.01220135935615766$ & $-2.104916128829678\times 10^{-5}$ \\ \hline
\multirow{3}{*}{$0.00035$} & a & $-0.01113604$ & $-8.9570\times 10^{-5}$ \\
& b & $-0.01114288854595917$ & $-9.327043407081445\times 10^{-5}$ \\
& c & $-0.01114288854595917$ & $-9.327043407081445\times 10^{-5}$ \\ \hline
\multirow{3}{*}{$0.00040$} & a & $-0.01008206$ & $-2.1402\times 10^{-4}$ \\
& b & $-0.01011729953739499$ & $-2.321591626792999\times 10^{-4}$ \\
& c & $-0.01011729953739499$ & $-2.321591626792999\times 10^{-4}$ \\ \hline
\multirow{3}{*}{$0.00045$} & a & $-0.00899479$ & $-3.7941\times 10^{-4}$ \\
& b & $-0.00909725070184054$ & $-4.263615594700631\times 10^{-4}$ \\
& c & $-0.00909725070184054$ & $-4.263615594700631\times 1^{-4}$ \\ \hline
\multirow{3}{*}{$0.00050$} & a & $-0.0078517$ & $-5.7415\times 10^{-4}$ \\
& b & $-0.00807076238659657$ & $-6.601708601710509\times 10^{-4}$ \\
& c & $-0.00807076238659657$ & $-6.601708601710509\times 10^{-4}$ \\ \hline
\end{tabular}

}

\end{table}

\begin{table}[H]
\caption{Resonances calculated by Kolosov\protect\cite{K87} (a) and by means
of the RPM with $D=30$ (b)}
\label{tab:n=10}%

\begin{subtable}[t]{\textwidth}
\caption*{$\Ket{9,0,0}$} \centering \scalebox{0.8}{
\begin{tabular}{|c|c|l|l|}
\hline \multicolumn{1}{|c|}{$10^5\,F$} &  &
\multicolumn{1}{c|}{Re$E$} &
\multicolumn{1}{c|}{Im$E$} \\
\hline
\multirow{2}{*}{2.0} & a & $-2.58557398\times 10^{-3}$ & $-9.509227\times 10^{-8}$\\
 & b & $-2.585573979364734\times 10^{-3}$ & $-9.509226978683682\times 10^{-8}$\\
 \hline
\multirow{2}{*}{3.0} & a & $-1.57105982\times 10^{-3}$ & $-3.959433\times 10^{-5}$\\
 & b & $-1.571059822031523\times 10^{-3}$ & $-3.959432995212450\times 10^{-5}$\\
 \hline
\multirow{2}{*}{4.0} & a & $-5.8496223\times 10^{-4}$ & $-1.6703408\times 10^{-4}$\\
 & b & $-5.849621042229387\times 10^{-4}$ & $-1.670340346132577\times 10^{-4}$\\
\hline
\end{tabular}

}

\end{subtable}
\begin{subtable}[t]{\textwidth}
\caption*{ $\Ket{0,0,9}$} \centering \scalebox{0.8}{
\begin{tabular}{|c|c|l|l|}
\hline \multicolumn{1}{|c|}{$10^5\,F$} &  &
\multicolumn{1}{c|}{Re$E$} &
\multicolumn{1}{c|}{Im$E$} \\
\hline
\multirow{2}{*}{2.0} & a & $-5.32440479\times 10^{-3}$ & $-3.9351431\times 10^{-5}$\\
 & b & $-5.324404794258087\times 10^{-3}$ & $-3.935143048784509\times 10^{-5}$\\
\hline
\multirow{2}{*}{3.0} & a & $-5.6483507\times 10^{-3}$ & $-3.263613\times 10^{-4}$\\
 & b & $-5.648350339949772\times 10^{-3}$ & $-3.263623549768137\times 10^{-4}$\\
\hline
\end{tabular}

}

\end{subtable}
\begin{subtable}[t]{\textwidth}
\caption*{ $\Ket{0,9,0}$ \label{tab:sb:0,9,0} } \centering
\scalebox{0.8}{
\begin{tabular}{|c|c|l|l|}
\hline \multicolumn{1}{|c|}{$10^5\,F$} &  &
\multicolumn{1}{c|}{Re$E$} &
\multicolumn{1}{c|}{Im$E$} \\
\hline
\multirow{2}{*}{1.4} & a & $-7.2120845\times 10^{-3}$ & $-4.0070215\times 10^{-5}$\\
 & b & $-7.212084472616482\times 10^{-3}$ & $-4.007021552503371\times 10^{-5}$\\
\hline
\multirow{2}{*}{1.8} & a & $-7.977367\times 10^{-3}$ & $-2.4391785\times 10^{-4}$\\
 & b & $-7.977367228278029\times 10^{-3}$ & $-2.439179991742931\times 10^{-4}$\\
\hline
\multirow{2}{*}{2.2} & a & $-8.660578\times 10^{-3}$ & $-5.32992\times 10^{-6}$\\
 & b & $-8.660579416493959\times 10^{-3}$ & $-5.329919686471733\times 10^{-4}$\\
\hline
\end{tabular}

}

\end{subtable}

\end{table}

\begin{table}[H]
\caption{Resonance $\Ket{39,0,0}$ calculated by means of the RPM with $D\leq
65$ }
\label{tab:n=40}\centering
\begin{tabular}{|c|l|l|}
\hline
\multicolumn{1}{|c|}{$10^7\,F$} & \multicolumn{1}{c|}{Re$E$} &
\multicolumn{1}{c|}{Im$E$} \\ \hline
1.0 & $-1.033131815036742\times 10^{-4}$ & $-1.410563208376528\times
10^{-12} $ \\
1.2 & $-6.597779434524293\times 10^{-4}$ & $-2.882918724695370\times 10^{-8}$
\\
1.4 & $-3.007626411426787\times 10^{-5}$ & $-8.858397640808244\times 10^{-7}$
\\
1.6 & $\phantom{-}5.730328096956075\times 10^{-6}$ & $-2.598217010741238\times 10^{-6}$
\\
1.8 & $\phantom{-}4.153413722356150\times 10^{-5}$ & $-4.277370190378267\times 10^{-6}$
\\
2.0 & $\phantom{-}7.701721282910781\times 10^{-5}$ & $-5.720318291587286\times 10^{-6}$
\\
2.2 & $\phantom{-}1.120415206875958\times 10^{-4}$ & $-6.975702947269457\times 10^{-6}$
\\ \hline
\end{tabular}
\end{table}

\begin{figure}[H]
\caption{Convergence of the RPM resonances for $F=0.001$}
\label{fig:conv1}
\begin{center}
\begin{subfigure}[b]{0.48\textwidth}
\resizebox{6cm}{!}{\input{conv1_Re}}
\end{subfigure}
\begin{subfigure}[b]{0.48\textwidth}
\resizebox{6cm}{!}{\input{conv1_Im}}
\end{subfigure}
\end{center}
\end{figure}
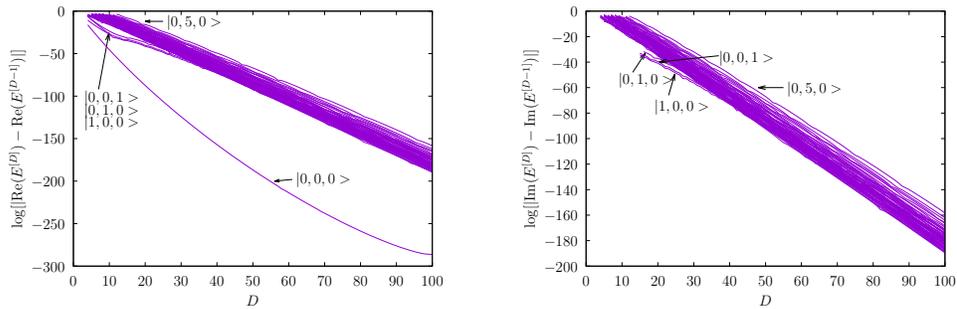

\end{document}

%% file: conv1_Re.tex
\begingroup
  \makeatletter
  \providecommand\color[2][]{%
    \GenericError{(gnuplot) \space\space\space\@spaces}{%
      Package color not loaded in conjunction with
      terminal option `colourtext'%
    }{See the gnuplot documentation for explanation.%
    }{Either use 'blacktext' in gnuplot or load the package
      color.sty in LaTeX.}%
    \renewcommand\color[2][]{}%
  }%
  \providecommand\includegraphics[2][]{%
    \GenericError{(gnuplot) \space\space\space\@spaces}{%
      Package graphicx or graphics not loaded%
    }{See the gnuplot documentation for explanation.%
    }{The gnuplot epslatex terminal needs graphicx.sty or graphics.sty.}%
    \renewcommand\includegraphics[2][]{}%
  }%
  \providecommand\rotatebox[2]{#2}%
  \@ifundefined{ifGPcolor}{%
    \newif\ifGPcolor
    \GPcolorfalse
  }{}%
  \@ifundefined{ifGPblacktext}{%
    \newif\ifGPblacktext
    \GPblacktexttrue
  }{}%
  \let\gplgaddtomacro\g@addto@macro
  \gdef\gplbacktext{}%
  \gdef\gplfronttext{}%
  \makeatother
  \ifGPblacktext
    \def\colorrgb#1{}%
    \def\colorgray#1{}%
  \else
    \ifGPcolor
      \def\colorrgb#1{\color[rgb]{#1}}%
      \def\colorgray#1{\color[gray]{#1}}%
      \expandafter\def\csname LTw\endcsname{\color{white}}%
      \expandafter\def\csname LTb\endcsname{\color{black}}%
      \expandafter\def\csname LTa\endcsname{\color{black}}%
      \expandafter\def\csname LT0\endcsname{\color[rgb]{1,0,0}}%
      \expandafter\def\csname LT1\endcsname{\color[rgb]{0,1,0}}%
      \expandafter\def\csname LT2\endcsname{\color[rgb]{0,0,1}}%
      \expandafter\def\csname LT3\endcsname{\color[rgb]{1,0,1}}%
      \expandafter\def\csname LT4\endcsname{\color[rgb]{0,1,1}}%
      \expandafter\def\csname LT5\endcsname{\color[rgb]{1,1,0}}%
      \expandafter\def\csname LT6\endcsname{\color[rgb]{0,0,0}}%
      \expandafter\def\csname LT7\endcsname{\color[rgb]{1,0.3,0}}%
      \expandafter\def\csname LT8\endcsname{\color[rgb]{0.5,0.5,0.5}}%
    \else
      \def\colorrgb#1{\color{black}}%
      \def\colorgray#1{\color[gray]{#1}}%
      \expandafter\def\csname LTw\endcsname{\color{white}}%
      \expandafter\def\csname LTb\endcsname{\color{black}}%
      \expandafter\def\csname LTa\endcsname{\color{black}}%
      \expandafter\def\csname LT0\endcsname{\color{black}}%
      \expandafter\def\csname LT1\endcsname{\color{black}}%
      \expandafter\def\csname LT2\endcsname{\color{black}}%
      \expandafter\def\csname LT3\endcsname{\color{black}}%
      \expandafter\def\csname LT4\endcsname{\color{black}}%
      \expandafter\def\csname LT5\endcsname{\color{black}}%
      \expandafter\def\csname LT6\endcsname{\color{black}}%
      \expandafter\def\csname LT7\endcsname{\color{black}}%
      \expandafter\def\csname LT8\endcsname{\color{black}}%
    \fi
  \fi
    \setlength{\unitlength}{0.0500bp}%
    \ifx\gptboxheight\undefined%
      \newlength{\gptboxheight}%
      \newlength{\gptboxwidth}%
      \newsavebox{\gptboxtext}%
    \fi%
    \setlength{\fboxrule}{0.5pt}%
    \setlength{\fboxsep}{1pt}%
\begin{picture}(7200.00,5040.00)%
    \gplgaddtomacro\gplbacktext{%
      \csname LTb\endcsname%
      \put(946,704){\makebox(0,0)[r]{\strut{}$-300$}}%
      \put(946,1383){\makebox(0,0)[r]{\strut{}$-250$}}%
      \put(946,2061){\makebox(0,0)[r]{\strut{}$-200$}}%
      \put(946,2740){\makebox(0,0)[r]{\strut{}$-150$}}%
      \put(946,3418){\makebox(0,0)[r]{\strut{}$-100$}}%
      \put(946,4097){\makebox(0,0)[r]{\strut{}$-50$}}%
      \put(946,4775){\makebox(0,0)[r]{\strut{}$0$}}%
      \put(1078,484){\makebox(0,0){\strut{}$0$}}%
      \put(1651,484){\makebox(0,0){\strut{}$10$}}%
      \put(2223,484){\makebox(0,0){\strut{}$20$}}%
      \put(2796,484){\makebox(0,0){\strut{}$30$}}%
      \put(3368,484){\makebox(0,0){\strut{}$40$}}%
      \put(3941,484){\makebox(0,0){\strut{}$50$}}%
      \put(4513,484){\makebox(0,0){\strut{}$60$}}%
      \put(5086,484){\makebox(0,0){\strut{}$70$}}%
      \put(5658,484){\makebox(0,0){\strut{}$80$}}%
      \put(6231,484){\makebox(0,0){\strut{}$90$}}%
      \put(6803,484){\makebox(0,0){\strut{}$100$}}%
      \put(4628,2115){\makebox(0,0)[l]{\strut{}$|0,0,0>$}}%
      \put(2567,4612){\makebox(0,0)[l]{\strut{}$|0,5,0>$}}%
      \put(1250,3418){\makebox(0,0)[l]{\strut{}$|0,0,1>$}}%
      \put(1250,3214){\makebox(0,0)[l]{\strut{}$|0,1,0>$}}%
      \put(1250,3011){\makebox(0,0)[l]{\strut{}$|1,0,0>$}}%
    }%
    \gplgaddtomacro\gplfronttext{%
      \csname LTb\endcsname%
      \put(176,2739){\rotatebox{-270}{\makebox(0,0){\strut{}$\mathrm{log}
      [|\mathrm{Re}(E^{[D]})-\mathrm{Re}(E^{[D-1]})|]$}}}
      \put(3940,154){\makebox(0,0){\strut{}$D$}}%
    }%
    \gplbacktext
    \put(0,0){\includegraphics{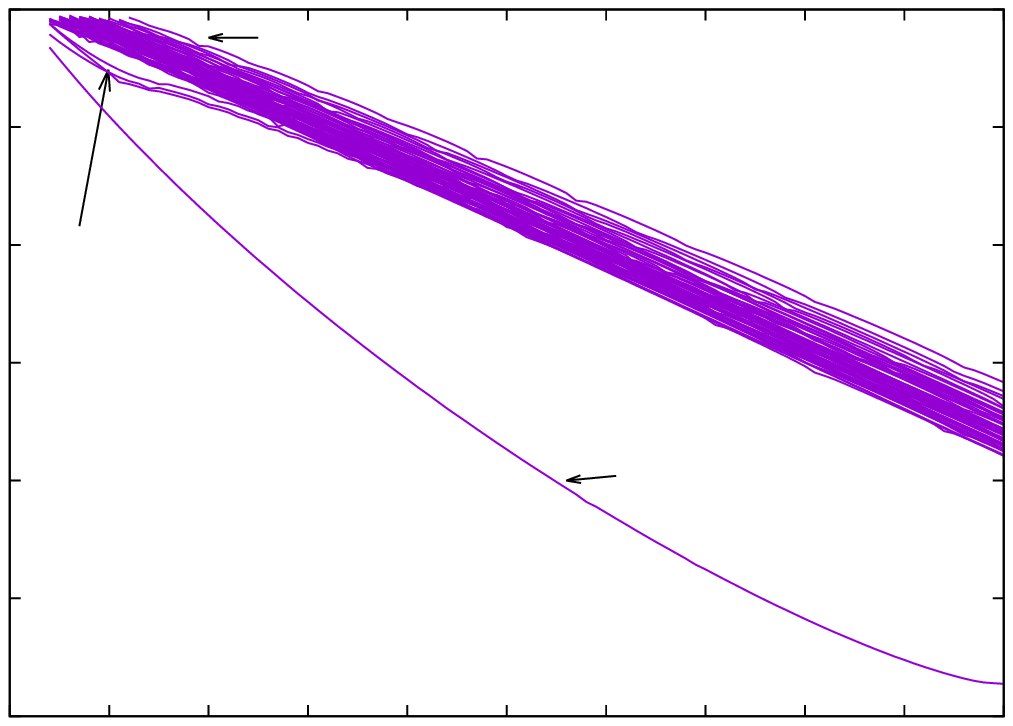}}%
    \gplfronttext
  \end{picture}%
\endgroup

%% file: conv1_Im.tex
\begingroup
  \makeatletter
  \providecommand\color[2][]{%
    \GenericError{(gnuplot) \space\space\space\@spaces}{%
      Package color not loaded in conjunction with
      terminal option `colourtext'%
    }{See the gnuplot documentation for explanation.%
    }{Either use 'blacktext' in gnuplot or load the package
      color.sty in LaTeX.}%
    \renewcommand\color[2][]{}%
  }%
  \providecommand\includegraphics[2][]{%
    \GenericError{(gnuplot) \space\space\space\@spaces}{%
      Package graphicx or graphics not loaded%
    }{See the gnuplot documentation for explanation.%
    }{The gnuplot epslatex terminal needs graphicx.sty or graphics.sty.}%
    \renewcommand\includegraphics[2][]{}%
  }%
  \providecommand\rotatebox[2]{#2}%
  \@ifundefined{ifGPcolor}{%
    \newif\ifGPcolor
    \GPcolorfalse
  }{}%
  \@ifundefined{ifGPblacktext}{%
    \newif\ifGPblacktext
    \GPblacktexttrue
  }{}%
  \let\gplgaddtomacro\g@addto@macro
  \gdef\gplbacktext{}%
  \gdef\gplfronttext{}%
  \makeatother
  \ifGPblacktext
    \def\colorrgb#1{}%
    \def\colorgray#1{}%
  \else
    \ifGPcolor
      \def\colorrgb#1{\color[rgb]{#1}}%
      \def\colorgray#1{\color[gray]{#1}}%
      \expandafter\def\csname LTw\endcsname{\color{white}}%
      \expandafter\def\csname LTb\endcsname{\color{black}}%
      \expandafter\def\csname LTa\endcsname{\color{black}}%
      \expandafter\def\csname LT0\endcsname{\color[rgb]{1,0,0}}%
      \expandafter\def\csname LT1\endcsname{\color[rgb]{0,1,0}}%
      \expandafter\def\csname LT2\endcsname{\color[rgb]{0,0,1}}%
      \expandafter\def\csname LT3\endcsname{\color[rgb]{1,0,1}}%
      \expandafter\def\csname LT4\endcsname{\color[rgb]{0,1,1}}%
      \expandafter\def\csname LT5\endcsname{\color[rgb]{1,1,0}}%
      \expandafter\def\csname LT6\endcsname{\color[rgb]{0,0,0}}%
      \expandafter\def\csname LT7\endcsname{\color[rgb]{1,0.3,0}}%
      \expandafter\def\csname LT8\endcsname{\color[rgb]{0.5,0.5,0.5}}%
    \else
      \def\colorrgb#1{\color{black}}%
      \def\colorgray#1{\color[gray]{#1}}%
      \expandafter\def\csname LTw\endcsname{\color{white}}%
      \expandafter\def\csname LTb\endcsname{\color{black}}%
      \expandafter\def\csname LTa\endcsname{\color{black}}%
      \expandafter\def\csname LT0\endcsname{\color{black}}%
      \expandafter\def\csname LT1\endcsname{\color{black}}%
      \expandafter\def\csname LT2\endcsname{\color{black}}%
      \expandafter\def\csname LT3\endcsname{\color{black}}%
      \expandafter\def\csname LT4\endcsname{\color{black}}%
      \expandafter\def\csname LT5\endcsname{\color{black}}%
      \expandafter\def\csname LT6\endcsname{\color{black}}%
      \expandafter\def\csname LT7\endcsname{\color{black}}%
      \expandafter\def\csname LT8\endcsname{\color{black}}%
    \fi
  \fi
    \setlength{\unitlength}{0.0500bp}%
    \ifx\gptboxheight\undefined%
      \newlength{\gptboxheight}%
      \newlength{\gptboxwidth}%
      \newsavebox{\gptboxtext}%
    \fi%
    \setlength{\fboxrule}{0.5pt}%
    \setlength{\fboxsep}{1pt}%
\begin{picture}(7200.00,5040.00)%
    \gplgaddtomacro\gplbacktext{%
      \csname LTb\endcsname%
      \put(946,704){\makebox(0,0)[r]{\strut{}$-200$}}%
      \put(946,1111){\makebox(0,0)[r]{\strut{}$-180$}}%
      \put(946,1518){\makebox(0,0)[r]{\strut{}$-160$}}%
      \put(946,1925){\makebox(0,0)[r]{\strut{}$-140$}}%
      \put(946,2332){\makebox(0,0)[r]{\strut{}$-120$}}%
      \put(946,2740){\makebox(0,0)[r]{\strut{}$-100$}}%
      \put(946,3147){\makebox(0,0)[r]{\strut{}$-80$}}%
      \put(946,3554){\makebox(0,0)[r]{\strut{}$-60$}}%
      \put(946,3961){\makebox(0,0)[r]{\strut{}$-40$}}%
      \put(946,4368){\makebox(0,0)[r]{\strut{}$-20$}}%
      \put(946,4775){\makebox(0,0)[r]{\strut{}$0$}}%
      \put(1078,484){\makebox(0,0){\strut{}$0$}}%
      \put(1651,484){\makebox(0,0){\strut{}$10$}}%
      \put(2223,484){\makebox(0,0){\strut{}$20$}}%
      \put(2796,484){\makebox(0,0){\strut{}$30$}}%
      \put(3368,484){\makebox(0,0){\strut{}$40$}}%
      \put(3941,484){\makebox(0,0){\strut{}$50$}}%
      \put(4513,484){\makebox(0,0){\strut{}$60$}}%
      \put(5086,484){\makebox(0,0){\strut{}$70$}}%
      \put(5658,484){\makebox(0,0){\strut{}$80$}}%
      \put(6231,484){\makebox(0,0){\strut{}$90$}}%
      \put(6803,484){\makebox(0,0){\strut{}$100$}}%
      \put(1593,3696){\makebox(0,0)[l]{\strut{}$|0,1,0>$}}%
      \put(2166,3269){\makebox(0,0)[l]{\strut{}$|1,0,0>$}}%
      \put(4284,3554){\makebox(0,0)[l]{\strut{}$|0,5,0>$}}%
      \put(3196,4042){\makebox(0,0)[l]{\strut{}$|0,0,1>$}}%
    }%
    \gplgaddtomacro\gplfronttext{%
      \csname LTb\endcsname%
      \put(176,2739){\rotatebox{-270}{\makebox(0,0){\strut{}$\mathrm{log}
      [|\mathrm{Im}(E^{[D]})-\mathrm{Im}(E^{[D-1]})|]$}}}
      \put(3940,154){\makebox(0,0){\strut{}$D$}}%
    }%
    \gplbacktext
    \put(0,0){\includegraphics{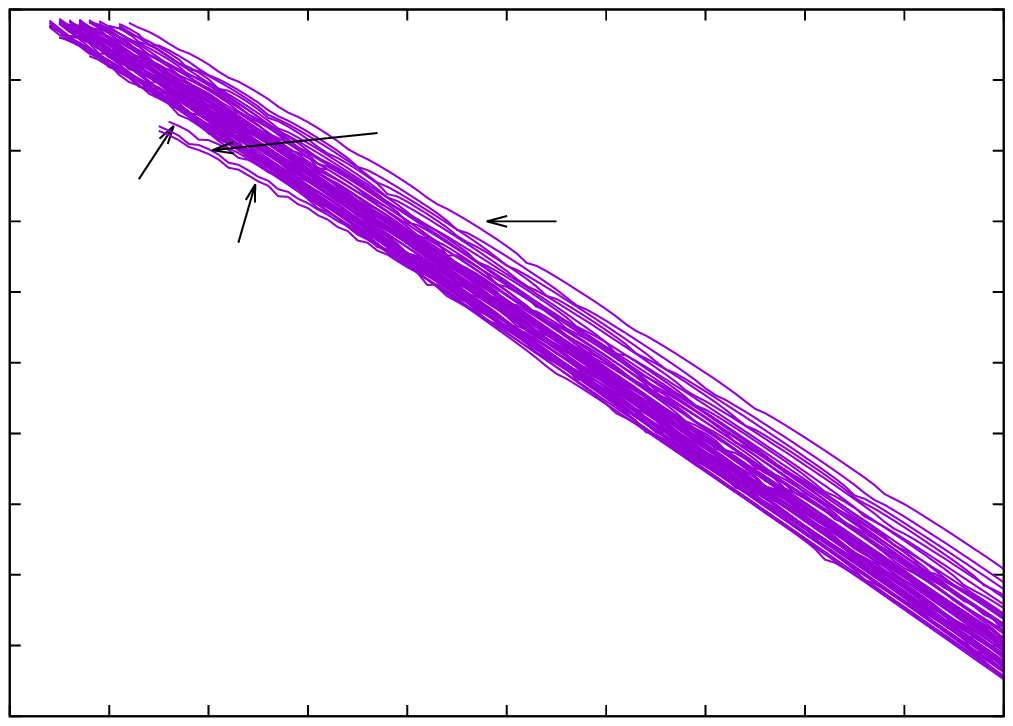}}%
    \gplfronttext
  \end{picture}%
\endgroup

%% file: conv2_Re.tex
\begingroup
  \makeatletter
  \providecommand\color[2][]{%
    \GenericError{(gnuplot) \space\space\space\@spaces}{%
      Package color not loaded in conjunction with
      terminal option `colourtext'%
    }{See the gnuplot documentation for explanation.%
    }{Either use 'blacktext' in gnuplot or load the package
      color.sty in LaTeX.}%
    \renewcommand\color[2][]{}%
  }%
  \providecommand\includegraphics[2][]{%
    \GenericError{(gnuplot) \space\space\space\@spaces}{%
      Package graphicx or graphics not loaded%
    }{See the gnuplot documentation for explanation.%
    }{The gnuplot epslatex terminal needs graphicx.sty or graphics.sty.}%
    \renewcommand\includegraphics[2][]{}%
  }%
  \providecommand\rotatebox[2]{#2}%
  \@ifundefined{ifGPcolor}{%
    \newif\ifGPcolor
    \GPcolorfalse
  }{}%
  \@ifundefined{ifGPblacktext}{%
    \newif\ifGPblacktext
    \GPblacktexttrue
  }{}%
  \let\gplgaddtomacro\g@addto@macro
  \gdef\gplbacktext{}%
  \gdef\gplfronttext{}%
  \makeatother
  \ifGPblacktext
    \def\colorrgb#1{}%
    \def\colorgray#1{}%
  \else
    \ifGPcolor
      \def\colorrgb#1{\color[rgb]{#1}}%
      \def\colorgray#1{\color[gray]{#1}}%
      \expandafter\def\csname LTw\endcsname{\color{white}}%
      \expandafter\def\csname LTb\endcsname{\color{black}}%
      \expandafter\def\csname LTa\endcsname{\color{black}}%
      \expandafter\def\csname LT0\endcsname{\color[rgb]{1,0,0}}%
      \expandafter\def\csname LT1\endcsname{\color[rgb]{0,1,0}}%
      \expandafter\def\csname LT2\endcsname{\color[rgb]{0,0,1}}%
      \expandafter\def\csname LT3\endcsname{\color[rgb]{1,0,1}}%
      \expandafter\def\csname LT4\endcsname{\color[rgb]{0,1,1}}%
      \expandafter\def\csname LT5\endcsname{\color[rgb]{1,1,0}}%
      \expandafter\def\csname LT6\endcsname{\color[rgb]{0,0,0}}%
      \expandafter\def\csname LT7\endcsname{\color[rgb]{1,0.3,0}}%
      \expandafter\def\csname LT8\endcsname{\color[rgb]{0.5,0.5,0.5}}%
    \else
      \def\colorrgb#1{\color{black}}%
      \def\colorgray#1{\color[gray]{#1}}%
      \expandafter\def\csname LTw\endcsname{\color{white}}%
      \expandafter\def\csname LTb\endcsname{\color{black}}%
      \expandafter\def\csname LTa\endcsname{\color{black}}%
      \expandafter\def\csname LT0\endcsname{\color{black}}%
      \expandafter\def\csname LT1\endcsname{\color{black}}%
      \expandafter\def\csname LT2\endcsname{\color{black}}%
      \expandafter\def\csname LT3\endcsname{\color{black}}%
      \expandafter\def\csname LT4\endcsname{\color{black}}%
      \expandafter\def\csname LT5\endcsname{\color{black}}%
      \expandafter\def\csname LT6\endcsname{\color{black}}%
      \expandafter\def\csname LT7\endcsname{\color{black}}%
      \expandafter\def\csname LT8\endcsname{\color{black}}%
    \fi
  \fi
    \setlength{\unitlength}{0.0500bp}%
    \ifx\gptboxheight\undefined%
      \newlength{\gptboxheight}%
      \newlength{\gptboxwidth}%
      \newsavebox{\gptboxtext}%
    \fi%
    \setlength{\fboxrule}{0.5pt}%
    \setlength{\fboxsep}{1pt}%
\begin{picture}(7200.00,5040.00)%
    \gplgaddtomacro\gplbacktext{%
      \csname LTb\endcsname%
      \put(946,704){\makebox(0,0)[r]{\strut{}$-200$}}%
      \put(946,1111){\makebox(0,0)[r]{\strut{}$-180$}}%
      \put(946,1518){\makebox(0,0)[r]{\strut{}$-160$}}%
      \put(946,1925){\makebox(0,0)[r]{\strut{}$-140$}}%
      \put(946,2332){\makebox(0,0)[r]{\strut{}$-120$}}%
      \put(946,2740){\makebox(0,0)[r]{\strut{}$-100$}}%
      \put(946,3147){\makebox(0,0)[r]{\strut{}$-80$}}%
      \put(946,3554){\makebox(0,0)[r]{\strut{}$-60$}}%
      \put(946,3961){\makebox(0,0)[r]{\strut{}$-40$}}%
      \put(946,4368){\makebox(0,0)[r]{\strut{}$-20$}}%
      \put(946,4775){\makebox(0,0)[r]{\strut{}$0$}}%
      \put(1078,484){\makebox(0,0){\strut{}$0$}}%
      \put(1651,484){\makebox(0,0){\strut{}$10$}}%
      \put(2223,484){\makebox(0,0){\strut{}$20$}}%
      \put(2796,484){\makebox(0,0){\strut{}$30$}}%
      \put(3368,484){\makebox(0,0){\strut{}$40$}}%
      \put(3941,484){\makebox(0,0){\strut{}$50$}}%
      \put(4513,484){\makebox(0,0){\strut{}$60$}}%
      \put(5086,484){\makebox(0,0){\strut{}$70$}}%
      \put(5658,484){\makebox(0,0){\strut{}$80$}}%
      \put(6231,484){\makebox(0,0){\strut{}$90$}}%
      \put(6803,484){\makebox(0,0){\strut{}$100$}}%
      \put(2509,2841){\makebox(0,0)[l]{\strut{}$|0,0,0>$}}%
      \put(4112,3981){\makebox(0,0)[l]{\strut{}$|0,5,0>$}}%
    }%
    \gplgaddtomacro\gplfronttext{%
      \csname LTb\endcsname%
      \put(176,2739){\rotatebox{-270}{\makebox(0,0){\strut{}$\mathrm{log}
      [|\mathrm{Re}(E^{[D]})-\mathrm{Re}(E^{[D-1]})|]$}}}
      \put(3940,154){\makebox(0,0){\strut{}$D$}}%
    }%
    \gplbacktext
    \put(0,0){\includegraphics{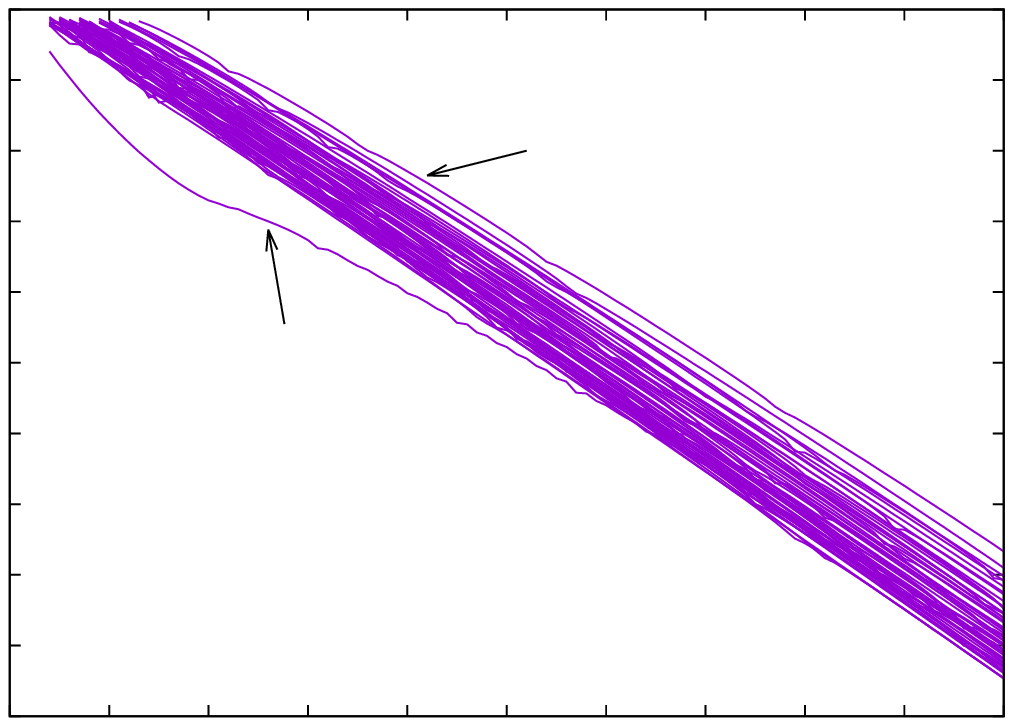}}%
    \gplfronttext
  \end{picture}%
\endgroup

%% file: conv2_Im.tex
\begingroup
  \makeatletter
  \providecommand\color[2][]{%
    \GenericError{(gnuplot) \space\space\space\@spaces}{%
      Package color not loaded in conjunction with
      terminal option `colourtext'%
    }{See the gnuplot documentation for explanation.%
    }{Either use 'blacktext' in gnuplot or load the package
      color.sty in LaTeX.}%
    \renewcommand\color[2][]{}%
  }%
  \providecommand\includegraphics[2][]{%
    \GenericError{(gnuplot) \space\space\space\@spaces}{%
      Package graphicx or graphics not loaded%
    }{See the gnuplot documentation for explanation.%
    }{The gnuplot epslatex terminal needs graphicx.sty or graphics.sty.}%
    \renewcommand\includegraphics[2][]{}%
  }%
  \providecommand\rotatebox[2]{#2}%
  \@ifundefined{ifGPcolor}{%
    \newif\ifGPcolor
    \GPcolorfalse
  }{}%
  \@ifundefined{ifGPblacktext}{%
    \newif\ifGPblacktext
    \GPblacktexttrue
  }{}%
  \let\gplgaddtomacro\g@addto@macro
  \gdef\gplbacktext{}%
  \gdef\gplfronttext{}%
  \makeatother
  \ifGPblacktext
    \def\colorrgb#1{}%
    \def\colorgray#1{}%
  \else
    \ifGPcolor
      \def\colorrgb#1{\color[rgb]{#1}}%
      \def\colorgray#1{\color[gray]{#1}}%
      \expandafter\def\csname LTw\endcsname{\color{white}}%
      \expandafter\def\csname LTb\endcsname{\color{black}}%
      \expandafter\def\csname LTa\endcsname{\color{black}}%
      \expandafter\def\csname LT0\endcsname{\color[rgb]{1,0,0}}%
      \expandafter\def\csname LT1\endcsname{\color[rgb]{0,1,0}}%
      \expandafter\def\csname LT2\endcsname{\color[rgb]{0,0,1}}%
      \expandafter\def\csname LT3\endcsname{\color[rgb]{1,0,1}}%
      \expandafter\def\csname LT4\endcsname{\color[rgb]{0,1,1}}%
      \expandafter\def\csname LT5\endcsname{\color[rgb]{1,1,0}}%
      \expandafter\def\csname LT6\endcsname{\color[rgb]{0,0,0}}%
      \expandafter\def\csname LT7\endcsname{\color[rgb]{1,0.3,0}}%
      \expandafter\def\csname LT8\endcsname{\color[rgb]{0.5,0.5,0.5}}%
    \else
      \def\colorrgb#1{\color{black}}%
      \def\colorgray#1{\color[gray]{#1}}%
      \expandafter\def\csname LTw\endcsname{\color{white}}%
      \expandafter\def\csname LTb\endcsname{\color{black}}%
      \expandafter\def\csname LTa\endcsname{\color{black}}%
      \expandafter\def\csname LT0\endcsname{\color{black}}%
      \expandafter\def\csname LT1\endcsname{\color{black}}%
      \expandafter\def\csname LT2\endcsname{\color{black}}%
      \expandafter\def\csname LT3\endcsname{\color{black}}%
      \expandafter\def\csname LT4\endcsname{\color{black}}%
      \expandafter\def\csname LT5\endcsname{\color{black}}%
      \expandafter\def\csname LT6\endcsname{\color{black}}%
      \expandafter\def\csname LT7\endcsname{\color{black}}%
      \expandafter\def\csname LT8\endcsname{\color{black}}%
    \fi
  \fi
    \setlength{\unitlength}{0.0500bp}%
    \ifx\gptboxheight\undefined%
      \newlength{\gptboxheight}%
      \newlength{\gptboxwidth}%
      \newsavebox{\gptboxtext}%
    \fi%
    \setlength{\fboxrule}{0.5pt}%
    \setlength{\fboxsep}{1pt}%
\begin{picture}(7200.00,5040.00)%
    \gplgaddtomacro\gplbacktext{%
      \csname LTb\endcsname%
      \put(946,704){\makebox(0,0)[r]{\strut{}$-200$}}%
      \put(946,1111){\makebox(0,0)[r]{\strut{}$-180$}}%
      \put(946,1518){\makebox(0,0)[r]{\strut{}$-160$}}%
      \put(946,1925){\makebox(0,0)[r]{\strut{}$-140$}}%
      \put(946,2332){\makebox(0,0)[r]{\strut{}$-120$}}%
      \put(946,2740){\makebox(0,0)[r]{\strut{}$-100$}}%
      \put(946,3147){\makebox(0,0)[r]{\strut{}$-80$}}%
      \put(946,3554){\makebox(0,0)[r]{\strut{}$-60$}}%
      \put(946,3961){\makebox(0,0)[r]{\strut{}$-40$}}%
      \put(946,4368){\makebox(0,0)[r]{\strut{}$-20$}}%
      \put(946,4775){\makebox(0,0)[r]{\strut{}$0$}}%
      \put(1078,484){\makebox(0,0){\strut{}$0$}}%
      \put(1651,484){\makebox(0,0){\strut{}$10$}}%
      \put(2223,484){\makebox(0,0){\strut{}$20$}}%
      \put(2796,484){\makebox(0,0){\strut{}$30$}}%
      \put(3368,484){\makebox(0,0){\strut{}$40$}}%
      \put(3941,484){\makebox(0,0){\strut{}$50$}}%
      \put(4513,484){\makebox(0,0){\strut{}$60$}}%
      \put(5086,484){\makebox(0,0){\strut{}$70$}}%
      \put(5658,484){\makebox(0,0){\strut{}$80$}}%
      \put(6231,484){\makebox(0,0){\strut{}$90$}}%
      \put(6803,484){\makebox(0,0){\strut{}$100$}}%
      \put(2166,2902){\makebox(0,0)[l]{\strut{}$|0,0,0>$}}%
      \put(4628,3452){\makebox(0,0)[l]{\strut{}$|0,5,0>$}}%
    }%
    \gplgaddtomacro\gplfronttext{%
      \csname LTb\endcsname%
      \put(176,2739){\rotatebox{-270}{\makebox(0,0){\strut{}$\mathrm{log}
      [|\mathrm{Im}(E^{[D]})-\mathrm{Im}(E^{[D-1]})|]$}}}
      \put(3940,154){\makebox(0,0){\strut{}$D$}}%
    }%
    \gplbacktext
    \put(0,0){\includegraphics{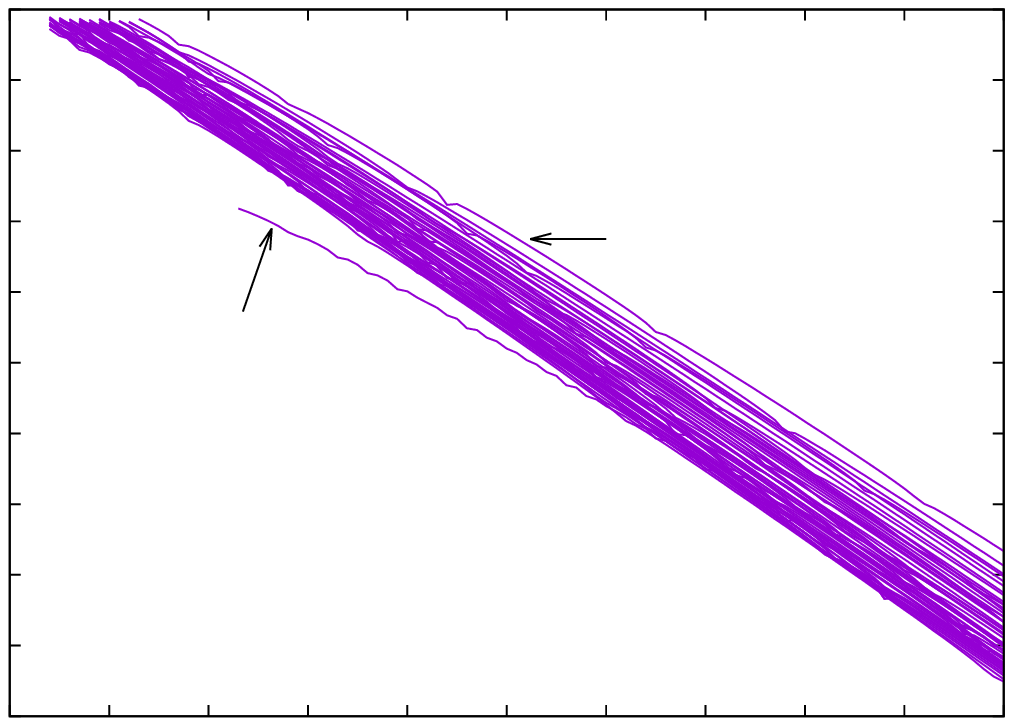}}%
    \gplfronttext
  \end{picture}%
\endgroup

%% file: RPM_vs_asinto.tex
\begingroup
  \makeatletter
  \providecommand\color[2][]{%
    \GenericError{(gnuplot) \space\space\space\@spaces}{%
      Package color not loaded in conjunction with
      terminal option `colourtext'%
    }{See the gnuplot documentation for explanation.%
    }{Either use 'blacktext' in gnuplot or load the package
      color.sty in LaTeX.}%
    \renewcommand\color[2][]{}%
  }%
  \providecommand\includegraphics[2][]{%
    \GenericError{(gnuplot) \space\space\space\@spaces}{%
      Package graphicx or graphics not loaded%
    }{See the gnuplot documentation for explanation.%
    }{The gnuplot epslatex terminal needs graphicx.sty or graphics.sty.}%
    \renewcommand\includegraphics[2][]{}%
  }%
  \providecommand\rotatebox[2]{#2}%
  \@ifundefined{ifGPcolor}{%
    \newif\ifGPcolor
    \GPcolorfalse
  }{}%
  \@ifundefined{ifGPblacktext}{%
    \newif\ifGPblacktext
    \GPblacktexttrue
  }{}%
  \let\gplgaddtomacro\g@addto@macro
  \gdef\gplbacktext{}%
  \gdef\gplfronttext{}%
  \makeatother
  \ifGPblacktext
    \def\colorrgb#1{}%
    \def\colorgray#1{}%
  \else
    \ifGPcolor
      \def\colorrgb#1{\color[rgb]{#1}}%
      \def\colorgray#1{\color[gray]{#1}}%
      \expandafter\def\csname LTw\endcsname{\color{white}}%
      \expandafter\def\csname LTb\endcsname{\color{black}}%
      \expandafter\def\csname LTa\endcsname{\color{black}}%
      \expandafter\def\csname LT0\endcsname{\color[rgb]{1,0,0}}%
      \expandafter\def\csname LT1\endcsname{\color[rgb]{0,1,0}}%
      \expandafter\def\csname LT2\endcsname{\color[rgb]{0,0,1}}%
      \expandafter\def\csname LT3\endcsname{\color[rgb]{1,0,1}}%
      \expandafter\def\csname LT4\endcsname{\color[rgb]{0,1,1}}%
      \expandafter\def\csname LT5\endcsname{\color[rgb]{1,1,0}}%
      \expandafter\def\csname LT6\endcsname{\color[rgb]{0,0,0}}%
      \expandafter\def\csname LT7\endcsname{\color[rgb]{1,0.3,0}}%
      \expandafter\def\csname LT8\endcsname{\color[rgb]{0.5,0.5,0.5}}%
    \else
      \def\colorrgb#1{\color{black}}%
      \def\colorgray#1{\color[gray]{#1}}%
      \expandafter\def\csname LTw\endcsname{\color{white}}%
      \expandafter\def\csname LTb\endcsname{\color{black}}%
      \expandafter\def\csname LTa\endcsname{\color{black}}%
      \expandafter\def\csname LT0\endcsname{\color{black}}%
      \expandafter\def\csname LT1\endcsname{\color{black}}%
      \expandafter\def\csname LT2\endcsname{\color{black}}%
      \expandafter\def\csname LT3\endcsname{\color{black}}%
      \expandafter\def\csname LT4\endcsname{\color{black}}%
      \expandafter\def\csname LT5\endcsname{\color{black}}%
      \expandafter\def\csname LT6\endcsname{\color{black}}%
      \expandafter\def\csname LT7\endcsname{\color{black}}%
      \expandafter\def\csname LT8\endcsname{\color{black}}%
    \fi
  \fi
    \setlength{\unitlength}{0.0500bp}%
    \ifx\gptboxheight\undefined%
      \newlength{\gptboxheight}%
      \newlength{\gptboxwidth}%
      \newsavebox{\gptboxtext}%
    \fi%
    \setlength{\fboxrule}{0.5pt}%
    \setlength{\fboxsep}{1pt}%
\begin{picture}(7200.00,5040.00)%
    \gplgaddtomacro\gplbacktext{%
      \csname LTb\endcsname%
      \put(946,704){\makebox(0,0)[r]{\strut{}-300}}%
      \put(946,1383){\makebox(0,0)[r]{\strut{}-250}}%
      \put(946,2061){\makebox(0,0)[r]{\strut{}-200}}%
      \put(946,2740){\makebox(0,0)[r]{\strut{}-150}}%
      \put(946,3418){\makebox(0,0)[r]{\strut{}-100}}%
      \put(946,4097){\makebox(0,0)[r]{\strut{}-50}}%
      \put(946,4775){\makebox(0,0)[r]{\strut{}0}}%
      \put(1078,484){\makebox(0,0){\strut{}$0$}}%
      \put(2032,484){\makebox(0,0){\strut{}$0.005$}}%
      \put(2986,484){\makebox(0,0){\strut{}$0.01$}}%
      \put(3941,484){\makebox(0,0){\strut{}$0.015$}}%
      \put(4895,484){\makebox(0,0){\strut{}$0.02$}}%
      \put(5849,484){\makebox(0,0){\strut{}$0.025$}}%
      \put(6803,484){\makebox(0,0){\strut{}$0.03$}}%
    }%
    \gplgaddtomacro\gplfronttext{%
      \csname LTb\endcsname%
      \put(176,2739){\rotatebox{-270}{\makebox(0,0){\strut{}$log_{10} [|Im(E)|]$}}}%
      \put(3940,154){\makebox(0,0){\strut{}$F$}}%
      \csname LTb\endcsname%
      \put(5816,1097){\makebox(0,0)[r]{\strut{}RPM}}%
      \csname LTb\endcsname%
      \put(5816,877){\makebox(0,0)[r]{\strut{}asymptotic}}%
    }%
    \gplbacktext
    \put(0,0){\includegraphics{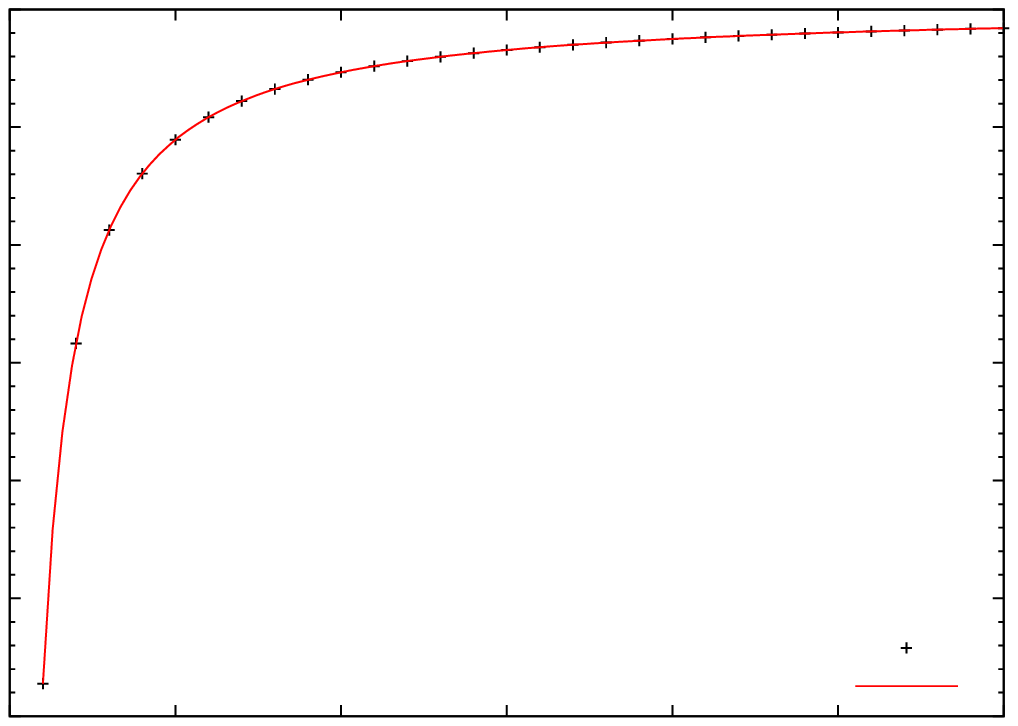}}%
    \gplfronttext
  \end{picture}%
\endgroup

%% file: 39-0-0_Re.tex
\begingroup
  \makeatletter
  \providecommand\color[2][]{%
    \GenericError{(gnuplot) \space\space\space\@spaces}{%
      Package color not loaded in conjunction with
      terminal option `colourtext'%
    }{See the gnuplot documentation for explanation.%
    }{Either use 'blacktext' in gnuplot or load the package
      color.sty in LaTeX.}%
    \renewcommand\color[2][]{}%
  }%
  \providecommand\includegraphics[2][]{%
    \GenericError{(gnuplot) \space\space\space\@spaces}{%
      Package graphicx or graphics not loaded%
    }{See the gnuplot documentation for explanation.%
    }{The gnuplot epslatex terminal needs graphicx.sty or graphics.sty.}%
    \renewcommand\includegraphics[2][]{}%
  }%
  \providecommand\rotatebox[2]{#2}%
  \@ifundefined{ifGPcolor}{%
    \newif\ifGPcolor
    \GPcolorfalse
  }{}%
  \@ifundefined{ifGPblacktext}{%
    \newif\ifGPblacktext
    \GPblacktexttrue
  }{}%
  \let\gplgaddtomacro\g@addto@macro
  \gdef\gplbacktext{}%
  \gdef\gplfronttext{}%
  \makeatother
  \ifGPblacktext
    \def\colorrgb#1{}%
    \def\colorgray#1{}%
  \else
    \ifGPcolor
      \def\colorrgb#1{\color[rgb]{#1}}%
      \def\colorgray#1{\color[gray]{#1}}%
      \expandafter\def\csname LTw\endcsname{\color{white}}%
      \expandafter\def\csname LTb\endcsname{\color{black}}%
      \expandafter\def\csname LTa\endcsname{\color{black}}%
      \expandafter\def\csname LT0\endcsname{\color[rgb]{1,0,0}}%
      \expandafter\def\csname LT1\endcsname{\color[rgb]{0,1,0}}%
      \expandafter\def\csname LT2\endcsname{\color[rgb]{0,0,1}}%
      \expandafter\def\csname LT3\endcsname{\color[rgb]{1,0,1}}%
      \expandafter\def\csname LT4\endcsname{\color[rgb]{0,1,1}}%
      \expandafter\def\csname LT5\endcsname{\color[rgb]{1,1,0}}%
      \expandafter\def\csname LT6\endcsname{\color[rgb]{0,0,0}}%
      \expandafter\def\csname LT7\endcsname{\color[rgb]{1,0.3,0}}%
      \expandafter\def\csname LT8\endcsname{\color[rgb]{0.5,0.5,0.5}}%
    \else
      \def\colorrgb#1{\color{black}}%
      \def\colorgray#1{\color[gray]{#1}}%
      \expandafter\def\csname LTw\endcsname{\color{white}}%
      \expandafter\def\csname LTb\endcsname{\color{black}}%
      \expandafter\def\csname LTa\endcsname{\color{black}}%
      \expandafter\def\csname LT0\endcsname{\color{black}}%
      \expandafter\def\csname LT1\endcsname{\color{black}}%
      \expandafter\def\csname LT2\endcsname{\color{black}}%
      \expandafter\def\csname LT3\endcsname{\color{black}}%
      \expandafter\def\csname LT4\endcsname{\color{black}}%
      \expandafter\def\csname LT5\endcsname{\color{black}}%
      \expandafter\def\csname LT6\endcsname{\color{black}}%
      \expandafter\def\csname LT7\endcsname{\color{black}}%
      \expandafter\def\csname LT8\endcsname{\color{black}}%
    \fi
  \fi
    \setlength{\unitlength}{0.0500bp}%
    \ifx\gptboxheight\undefined%
      \newlength{\gptboxheight}%
      \newlength{\gptboxwidth}%
      \newsavebox{\gptboxtext}%
    \fi%
    \setlength{\fboxrule}{0.5pt}%
    \setlength{\fboxsep}{1pt}%
\begin{picture}(7200.00,5040.00)%
    \gplgaddtomacro\gplbacktext{%
      \csname LTb\endcsname%
      \put(814,704){\makebox(0,0)[r]{\strut{}$-15$}}%
      \put(814,1383){\makebox(0,0)[r]{\strut{}$-10$}}%
      \put(814,2061){\makebox(0,0)[r]{\strut{}$-5$}}%
      \put(814,2740){\makebox(0,0)[r]{\strut{}$0$}}%
      \put(814,3418){\makebox(0,0)[r]{\strut{}$5$}}%
      \put(814,4097){\makebox(0,0)[r]{\strut{}$10$}}%
      \put(814,4775){\makebox(0,0)[r]{\strut{}$15$}}%
      \put(1364,484){\makebox(0,0){\strut{}$1$}}%
      \put(2201,484){\makebox(0,0){\strut{}$1.2$}}%
      \put(3038,484){\makebox(0,0){\strut{}$1.4$}}%
      \put(3874,484){\makebox(0,0){\strut{}$1.6$}}%
      \put(4711,484){\makebox(0,0){\strut{}$1.8$}}%
      \put(5548,484){\makebox(0,0){\strut{}$2$}}%
      \put(6385,484){\makebox(0,0){\strut{}$2.2$}}%
    }%
    \gplgaddtomacro\gplfronttext{%
      \csname LTb\endcsname%
      \put(176,2739){\rotatebox{-270}{\makebox(0,0){\strut{}$10^5\,Re(E)$}}}%
      \put(3874,154){\makebox(0,0){\strut{}$10^5\,F$}}%
      \csname LTb\endcsname%
      \put(1210,4602){\makebox(0,0)[r]{\strut{}a}}%
      \csname LTb\endcsname%
      \put(1210,4382){\makebox(0,0)[r]{\strut{}b}}%
    }%
    \gplbacktext
    \put(0,0){\includegraphics{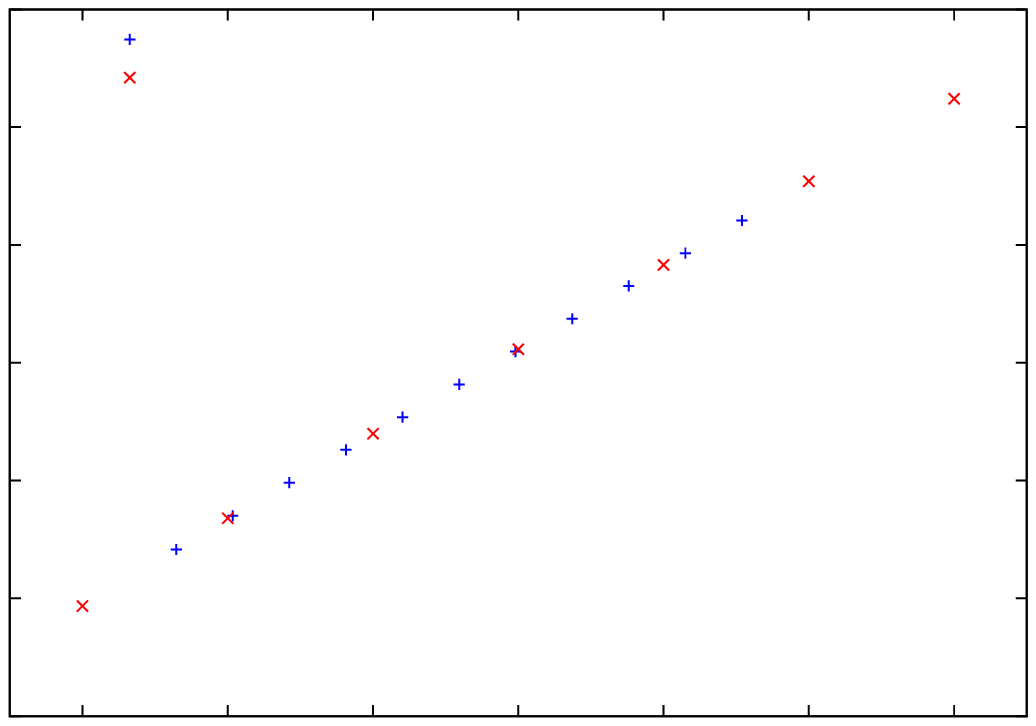}}%
    \gplfronttext
  \end{picture}%
\endgroup

%% file: 39-0-0_Im.tex
\begingroup
  \makeatletter
  \providecommand\color[2][]{%
    \GenericError{(gnuplot) \space\space\space\@spaces}{%
      Package color not loaded in conjunction with
      terminal option `colourtext'%
    }{See the gnuplot documentation for explanation.%
    }{Either use 'blacktext' in gnuplot or load the package
      color.sty in LaTeX.}%
    \renewcommand\color[2][]{}%
  }%
  \providecommand\includegraphics[2][]{%
    \GenericError{(gnuplot) \space\space\space\@spaces}{%
      Package graphicx or graphics not loaded%
    }{See the gnuplot documentation for explanation.%
    }{The gnuplot epslatex terminal needs graphicx.sty or graphics.sty.}%
    \renewcommand\includegraphics[2][]{}%
  }%
  \providecommand\rotatebox[2]{#2}%
  \@ifundefined{ifGPcolor}{%
    \newif\ifGPcolor
    \GPcolorfalse
  }{}%
  \@ifundefined{ifGPblacktext}{%
    \newif\ifGPblacktext
    \GPblacktexttrue
  }{}%
  \let\gplgaddtomacro\g@addto@macro
  \gdef\gplbacktext{}%
  \gdef\gplfronttext{}%
  \makeatother
  \ifGPblacktext
    \def\colorrgb#1{}%
    \def\colorgray#1{}%
  \else
    \ifGPcolor
      \def\colorrgb#1{\color[rgb]{#1}}%
      \def\colorgray#1{\color[gray]{#1}}%
      \expandafter\def\csname LTw\endcsname{\color{white}}%
      \expandafter\def\csname LTb\endcsname{\color{black}}%
      \expandafter\def\csname LTa\endcsname{\color{black}}%
      \expandafter\def\csname LT0\endcsname{\color[rgb]{1,0,0}}%
      \expandafter\def\csname LT1\endcsname{\color[rgb]{0,1,0}}%
      \expandafter\def\csname LT2\endcsname{\color[rgb]{0,0,1}}%
      \expandafter\def\csname LT3\endcsname{\color[rgb]{1,0,1}}%
      \expandafter\def\csname LT4\endcsname{\color[rgb]{0,1,1}}%
      \expandafter\def\csname LT5\endcsname{\color[rgb]{1,1,0}}%
      \expandafter\def\csname LT6\endcsname{\color[rgb]{0,0,0}}%
      \expandafter\def\csname LT7\endcsname{\color[rgb]{1,0.3,0}}%
      \expandafter\def\csname LT8\endcsname{\color[rgb]{0.5,0.5,0.5}}%
    \else
      \def\colorrgb#1{\color{black}}%
      \def\colorgray#1{\color[gray]{#1}}%
      \expandafter\def\csname LTw\endcsname{\color{white}}%
      \expandafter\def\csname LTb\endcsname{\color{black}}%
      \expandafter\def\csname LTa\endcsname{\color{black}}%
      \expandafter\def\csname LT0\endcsname{\color{black}}%
      \expandafter\def\csname LT1\endcsname{\color{black}}%
      \expandafter\def\csname LT2\endcsname{\color{black}}%
      \expandafter\def\csname LT3\endcsname{\color{black}}%
      \expandafter\def\csname LT4\endcsname{\color{black}}%
      \expandafter\def\csname LT5\endcsname{\color{black}}%
      \expandafter\def\csname LT6\endcsname{\color{black}}%
      \expandafter\def\csname LT7\endcsname{\color{black}}%
      \expandafter\def\csname LT8\endcsname{\color{black}}%
    \fi
  \fi
    \setlength{\unitlength}{0.0500bp}%
    \ifx\gptboxheight\undefined%
      \newlength{\gptboxheight}%
      \newlength{\gptboxwidth}%
      \newsavebox{\gptboxtext}%
    \fi%
    \setlength{\fboxrule}{0.5pt}%
    \setlength{\fboxsep}{1pt}%
\begin{picture}(7200.00,5040.00)%
    \gplgaddtomacro\gplbacktext{%
      \csname LTb\endcsname%
      \put(682,814){\makebox(0,0)[r]{\strut{}$-7$}}%
      \put(682,1364){\makebox(0,0)[r]{\strut{}$-6$}}%
      \put(682,1914){\makebox(0,0)[r]{\strut{}$-5$}}%
      \put(682,2464){\makebox(0,0)[r]{\strut{}$-4$}}%
      \put(682,3015){\makebox(0,0)[r]{\strut{}$-3$}}%
      \put(682,3565){\makebox(0,0)[r]{\strut{}$-2$}}%
      \put(682,4115){\makebox(0,0)[r]{\strut{}$-1$}}%
      \put(682,4665){\makebox(0,0)[r]{\strut{}$0$}}%
      \put(1242,484){\makebox(0,0){\strut{}$1$}}%
      \put(2097,484){\makebox(0,0){\strut{}$1.2$}}%
      \put(2953,484){\makebox(0,0){\strut{}$1.4$}}%
      \put(3808,484){\makebox(0,0){\strut{}$1.6$}}%
      \put(4664,484){\makebox(0,0){\strut{}$1.8$}}%
      \put(5520,484){\makebox(0,0){\strut{}$2$}}%
      \put(6375,484){\makebox(0,0){\strut{}$2.2$}}%
    }%
    \gplgaddtomacro\gplfronttext{%
      \csname LTb\endcsname%
      \put(176,2739){\rotatebox{-270}{\makebox(0,0){\strut{}$10^6\,Im(E)$}}}%
      \put(3808,154){\makebox(0,0){\strut{}$10^5\,F$}}%
      \csname LTb\endcsname%
      \put(5816,4602){\makebox(0,0)[r]{\strut{}a}}%
      \csname LTb\endcsname%
      \put(5816,4382){\makebox(0,0)[r]{\strut{}b}}%
    }%
    \gplbacktext
    \put(0,0){\includegraphics{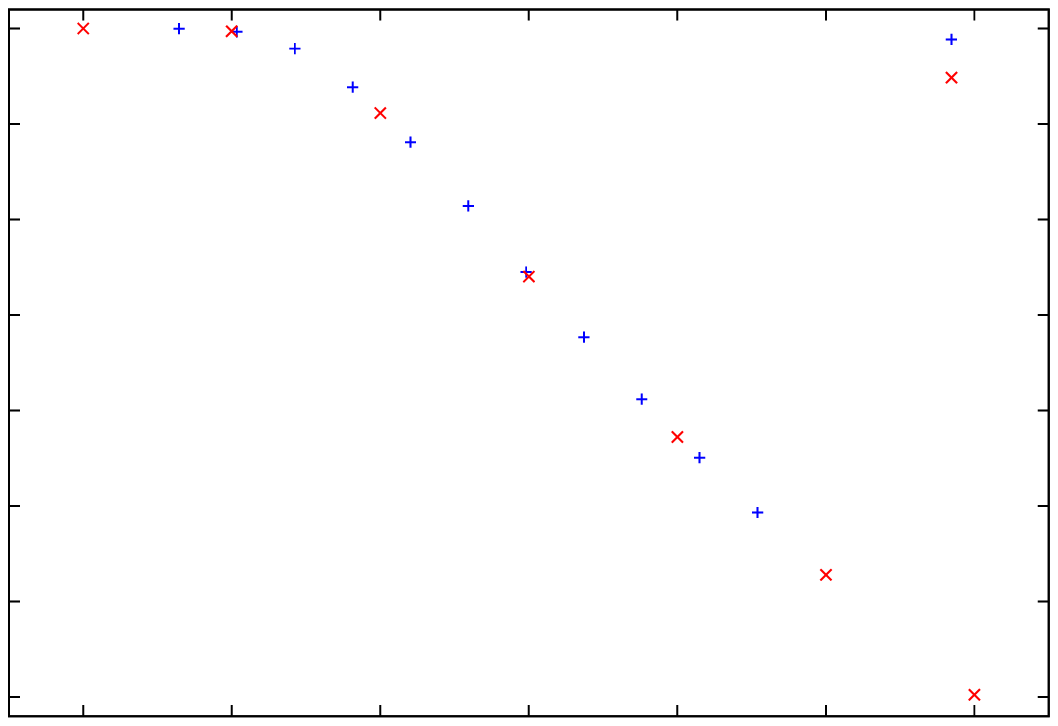}}%
    \gplfronttext
  \end{picture}%
\endgroup